\pdfoutput=1
\documentclass[%
 preprint,
 amsmath,amssymb,
 aps,
prl
]{revtex4-2}

\usepackage[english]{babel}
\usepackage{graphicx}
\usepackage{dcolumn}
\usepackage{bm}
\usepackage{nicefrac}
\usepackage{hyperref}

\usepackage{soul}
\usepackage[usernames,dvipsnames]{xcolor}
\usepackage{tikz}
\usepackage{comment}

\usepackage[protrusion=false,expansion=true,kerning=true,babel=true,final]{microtype}

\def\doubleunderline#1{\boldsymbol{{#1}}}

\DeclareMathOperator{\tr}{tr}

\graphicspath{{Figures/}}

\begin{document}

\preprint{APS/123-QED}

\title{Soft kirigami composites for form-finding of fully flexible deployables}


\author{Jan Zavodnik$^{1}$}
\author{Yunbo Wang$^{2}$}
\author{Wenzhong Yan$^{2}$}
\author{Miha Brojan$^{2}$}\email{M.B.: miha.brojan@fs.uni-lj.si}
\author{M. K. Jawed$^{1}$}\email{M.K.J.: khalidjm@seas.ucla.edu}

\affiliation{\footnotesize 
$^1$University of Ljubljana, Faculty of Mechanical Engineering, Laboratory for Nonlinear Mechanics, A\v{s}ker\v{c}eva 6, SI-1000, Ljubljana, Slovenia\\
$^2$University of California, Los Angeles, Department of Mechanical and Aerospace Engineering, 420 Westwood Plaza, Los Angeles, CA, USA 90024
}

\begin{abstract}
We introduce a new class of thin flexible structures that morph from a flat shape into prescribed 3D shapes without an external stimulus such as mechanical loads or heat. To achieve control over the target shape, two different concepts are coupled. First, motivated by biological growth, strain mismatch is applied between the flat composite layers to transform it into a 3D shape. Depending on the amount of the applied strain mismatch, the transformation involves buckling into one of the available finite number of mode shapes. Second, inspired by kirigami, portions of the material are removed from one of the layers according to a specific pattern. This dramatically increases the number of possible 3D shapes and allows us to attain specific topologies. We devise an experimental apparatus that allows precise control of the strain mismatch. An inverse problem is posed, where starting from a given target shape, the physical parameters that make these shapes possible are determined. To show how the concept works, we focus on circular composite plates and design a kirigami pattern that yields a hemispherical structure. Our analysis combines a theoretical approach with numerical simulations and physical experiments to understand and predict the transition from 2D to 3D shapes. The tools developed here can be extended to attain arbitrary 3D shapes. The initially flat shape suggests that conventional additive manufacturing techniques can be used to functionalize the soft kirigami composites to fabricate, for example, deployable 3D shapes, smart skins, and soft electromagnetic metasurfaces.
\end{abstract}

\pacs{Valid PACS appear here}

\maketitle

\begin{figure}[b!]
    \centering
    \includegraphics[width=0.98\textwidth]{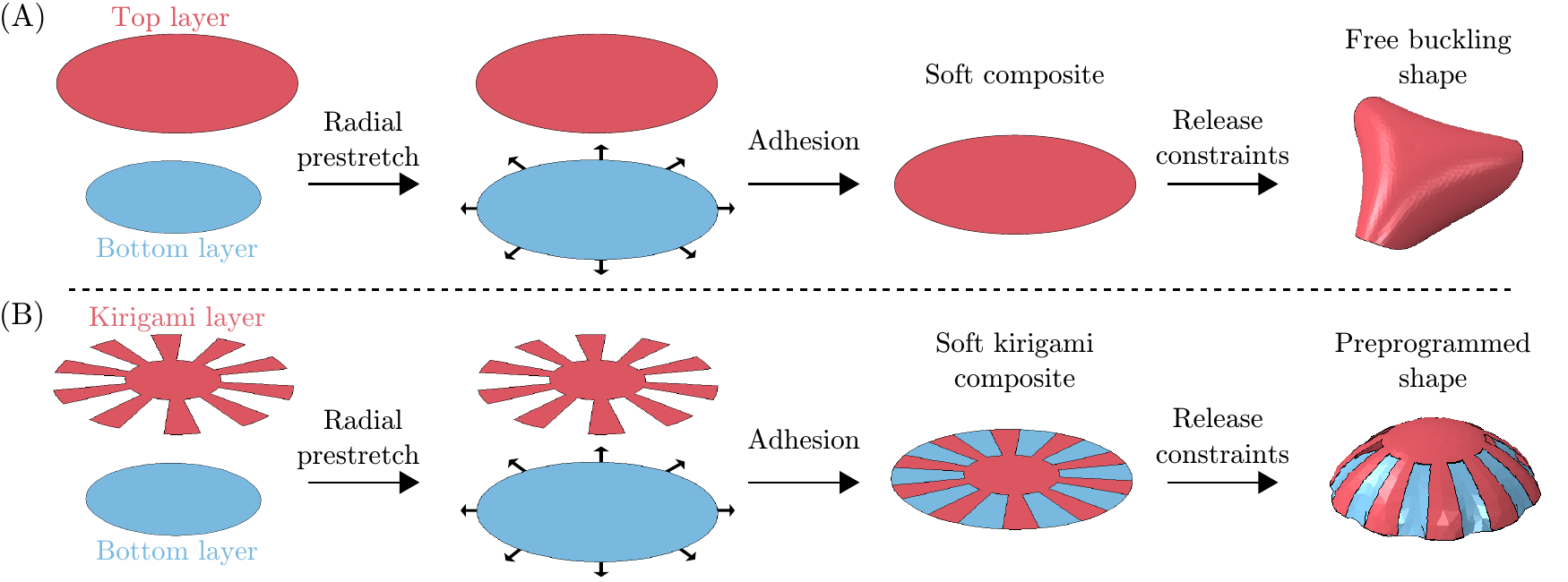}
    \caption{
    Overview of the problem. 
    (A) Of two circular layers, the bottom layer is stretched and adhered to the unstretched top layer to obtain a soft composite. Upon releasing the constraints, the strain mismatch between the two layers induces symmetry breaking and the composite may assume a buckled shape with a finite number ($3$ in this case) of waves along the circumference.
    (B) Replacing the top circular layer with a kirigami layer (layer with cuts) can lead to a preprogrammed shape via the same procedure. The shape is governed by the kirigami pattern. In this case, the goal was to obtain an axisymmetrical cap.
    } 
    \label{fig:overview}
\end{figure}


Morphing planar structures into preprogrammed three-dimensional shapes has applications in engineering across a wide range of length-scales from microns~\cite{xu2015assembly} to meters~\cite{panetta2019x}. Since conventional additive and subtractive manufacturing techniques typically support planar fabrication, morphing from 2D to 3D is a promising way to achieve 3D functional surfaces for use in, e.g., curvy electronics for wearables~\cite{sim2019three}, camouflaging~\cite{pikul2017stretchable}, structural health monitoring~\cite{wang2022recent}, multi-functional soft machines~\cite{ding2020multifunctional}, with existing manufacturing tools. Since the introduction of bimetallic strips~\cite{betts1993john} in the 18th century, morphing of slender structures into desired shapes has been actuated via heat~\cite{conti2002soft, saed2019molecularly}, light~\cite{wie2016photomotility}, electromagnetism~\cite{ni2022soft}, chemical gradient~\cite{shim2015dynamic}, growth~\cite{ambrosi2019growth}, and various forms of external stimuli. Another approach to morphing is inspired by origami, which has been particularly successful in deployable aerospace structures~\cite{dureisseix2012overview} that require small storage space but a large surface area. Such deployables are typically piecewise rigid and may require multiple springs, support structures, or other mechanisms for deployment. In this paper, a new class of deployables are envisioned that spontaneously morph from a planar shape to a prescribed 3D shape and are fully flexible without any rigid parts.

Towards fully flexible shells that morph from a 2D shape to a prescribed 3D topology, previous works on imposing mechanical loads and boundary conditions on sheets provide a solid foundation. Simply pulling a thin elastic sheet can induce 3D deformation through wrinkling instabilities~\cite{davidovitch2011prototypical}. Grason et al. showed that imposing curvature to elastic sheets leads to distinct types of structural instabilities~\cite{grason2013universal}. We consider this as an {\em incompatibility} of topologies leading to complex shapes. A simple instantiation of this concept can be achieved by draping a flat cloth around a spherical object. If the radius of the sphere is much larger than the size of the cloth, the planar cloth may assume the imposed 3D shape. However, if the sizes of the cloth and the sphere are comparable, wrinkles and crumples form. Geometric frustration leading to mechanical instabilities in sheets manifests itself in geometrically incompatible confinement of solids. Davidovitch et al.~\cite{davidovitch2019geometrically} studied a class of such problems in which the topography imposed on a thin solid body is incompatible with its intrinsic metric and wrinkles emerged as a consequence. These earlier works on the type and size of patterns during 2D to 3D transition inspired our solution to obtain 3D shapes with a prescribed target metric. Specifically, to avoid wrinkles (which depend on geometry and material stiffness) on the 3D shape, we explore the removal of the material to relieve geometric frustration and study its dependence on geometric and material properties.

New possibilities for shape selection open up when strain mismatch is introduced within a flat structure~\cite{pezzulla2015morphing, pezzulla2016geometry, pezzulla2017curvature, pezzulla2018curvature, desimone2018spontaneous}. 
Pezzulla et al.~\cite{pezzulla2018curvature} studied geometric frustration between multiple parts of the body that leads to 3D deformation of the naturally planar object. Strain mismatch introduced into any part of a thin body by, e.g., heating, growth, or swelling, can drastically affect the morphology of the entire object and induce mechanical instabilities. Such morphological changes are preponderant in biological structures and are often necessary for their functionality; examples include Venus flytrap, growing leaves, and the writhing of tendril bearing climbers in plants and formation of brains, lungs, and guts in animals.
Combining geometric confinement with strain mismatch can open pathways to an even broader class of shapes. Stein et al.~\cite{stein2019buckling} used residual swelling and geometric confinement to generate a range of shapes including saddles, rolled sheets, cylinders, and spherical sections.
Our work synergistically combines emergence of mechanical instabilities and strain mismatch to fabricate composite shells that morph from a flat shape to a prescribed 3D shape.

In this paper, we emphasize having control over the final 3D shape that morphs from an initially planar shape in contrast with prior works that had access to only a finite number of shapes. Our work uses a hemispherical smooth shape as a representative 3D shape. An intuitive (but ultimately incorrect) approach is presented in Fig.~\ref{fig:overview}A, where a ``bottom'' layer is radially stretched and attached to a ``top'' unstretched layer to form a soft composite shell. The shape of this composite is 3D but not hemispherical. A number of distinct shapes can emerge in this system, which are analyzed by F\"{o}ppl-Von K\'{a}rm\'{a}n plate theory, numerical simulations using the finite element method, and physical experiments. This lays the foundation for a solution to the hemispherical problem. Kirigami (i.e., removal of material) to relieve geometric frustration is introduced; a combination of kirigami and strain mismatch is proposed to access arbitrary 3D shapes. The interplay between the kirigami pattern, strain mismatch and the final shape is explored through experiments, simulations, and theoretical analysis. Fig.~\ref{fig:overview}B shows an example where a hemispherical shape is achieved by tuning the kirigami pattern and the strain mismatch. The concept of combining kirigami and strain mismatch can now be generalized to achieve arbitrary 3D shapes beyond just the hemispherical ones. A new experimental apparatus is designed and fabricated to impose uniform biaxial stretch onto the bottom layer while avoiding wrinkling due to Poisson\rq{}s effect~\cite{cerda2002wrinkling}.

Our paper is organized as follows. We commence with a description of the experimental apparatus in {\em Physical Experiments}. Various shapes resulting from two circular layers with strain mismatch is described in {\em Soft Circular Composites}. Kirigami on this composite introduced in {\em Soft Kirigami Composites} and methods to achieve the target hemispherical shape are discussed.

\begin{figure}[h!]
\centering
    \includegraphics[width=0.45\columnwidth]{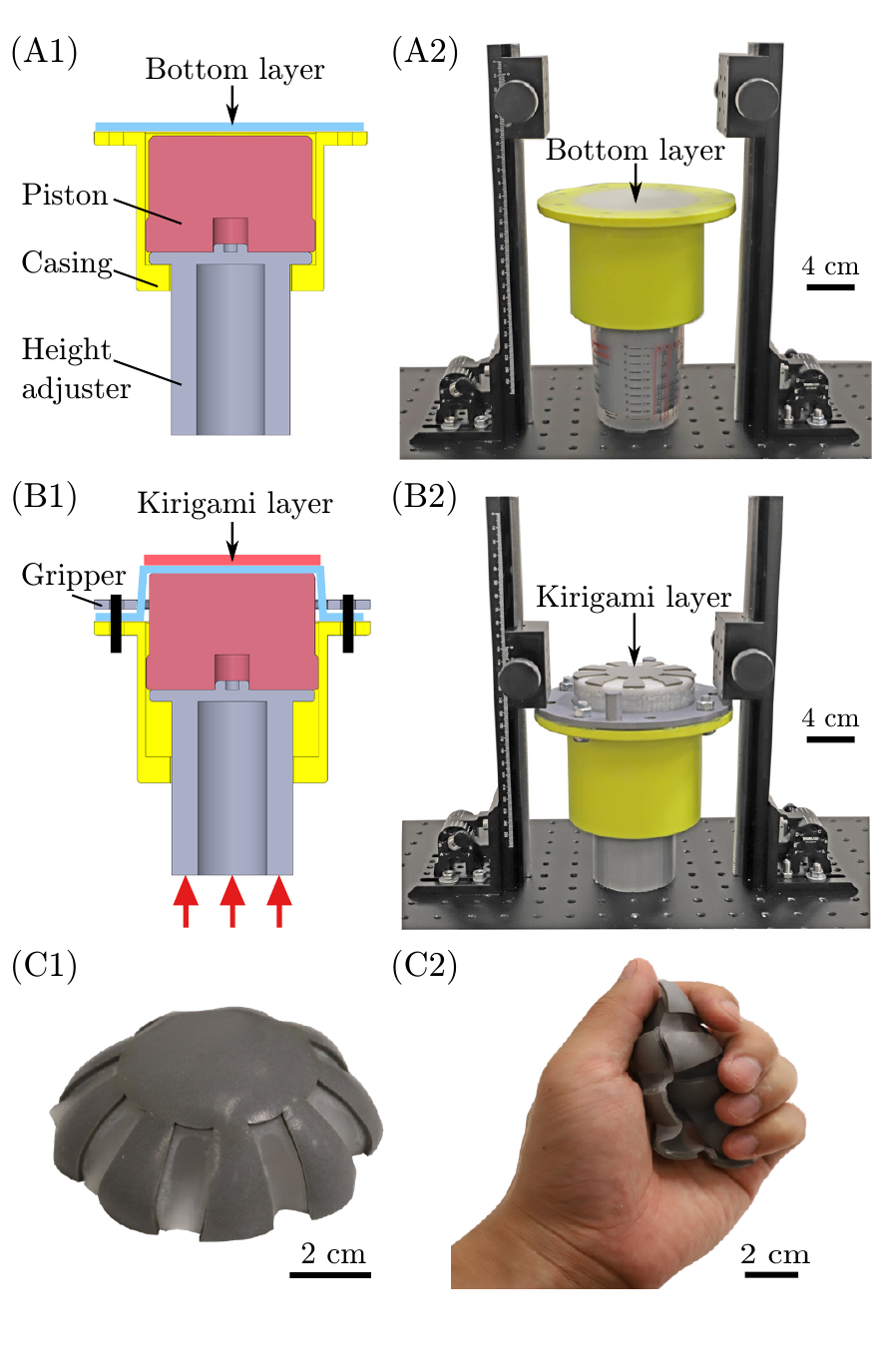}
    \caption{
    Experimental system. (A1) Schematic representation of the experimental system comprised of piston, casing, height adjuster, and annulus-shaped gripper. The bottom layer is attached above the piston.
    (A2) Snapshot of the system when bottom layer is still unstretched.
    (B1) Bottom layer is stretched by raising the piston and the kirigami layer is adhered on top.
    (B2) Snapshot of the system with soft kirigami composite still under constraints.
    (C1) After the constraints are removed by cutting the bottom layer from the gripper, the soft kirigami composite assumes the shape of a (quasi)axisymmetrical cap.
    (C2) The resulting composite is fully (elastically) flexible. It can be orderly folded or crumpled and thus stored in confined spaces, yet it returns to its preprogrammed shape upon the release of constraints.
    }
\label{fig:exptSetup}
\end{figure}

\section*{Physical Experiments}

The photograph of our custom designed experimental apparatus is presented in Fig.~\ref{fig:exptSetup}A1-A2. Two main components -- a 3D-printed cylinder assembly and two vertically positioned linear translation stages -- make up the set. The hyperelastic materials of the bottom layer (VHB 4910, 3M) and the top layer (VHB 4950, 3M) have double-side stickiness. Material properties are listed in \textbf{Materials and Methods}. As shown in Fig.~\ref{fig:exptSetup}A2, the bottom layer is placed on top of the cylindrical casing. The contacting part is pressed by a 3d-printed gripper in in Fig.~\ref{fig:exptSetup}B1-B2. 
%
The assembly is then placed between two vertical linear translation stages and the stages are moved to a specific height corresponding to a speicific prestretch. As the stages are moved upward, the piston -- pushed by the height adjuster -- also moves up but the outer edge of the bottom layer remains gripped onto the cylindrical casing. This leads to a uniform biaxial stretch in the hyperelastic material. The relation between the imposed stretch and height can be determined from simple geometry. The top layer is glued onto the stretched bottom layer to form a bilayer composite. The bilayer composite is then cut out from the apparatus and released from the gripper. This resulting 3D structure is referred to as the ``soft kirigami composite" in this paper.

\section*{Soft Circular Composites}\label{sec:SoftCircularComposites}

\begin{figure}[h!]
    \centering
    \includegraphics[width=0.98\textwidth]{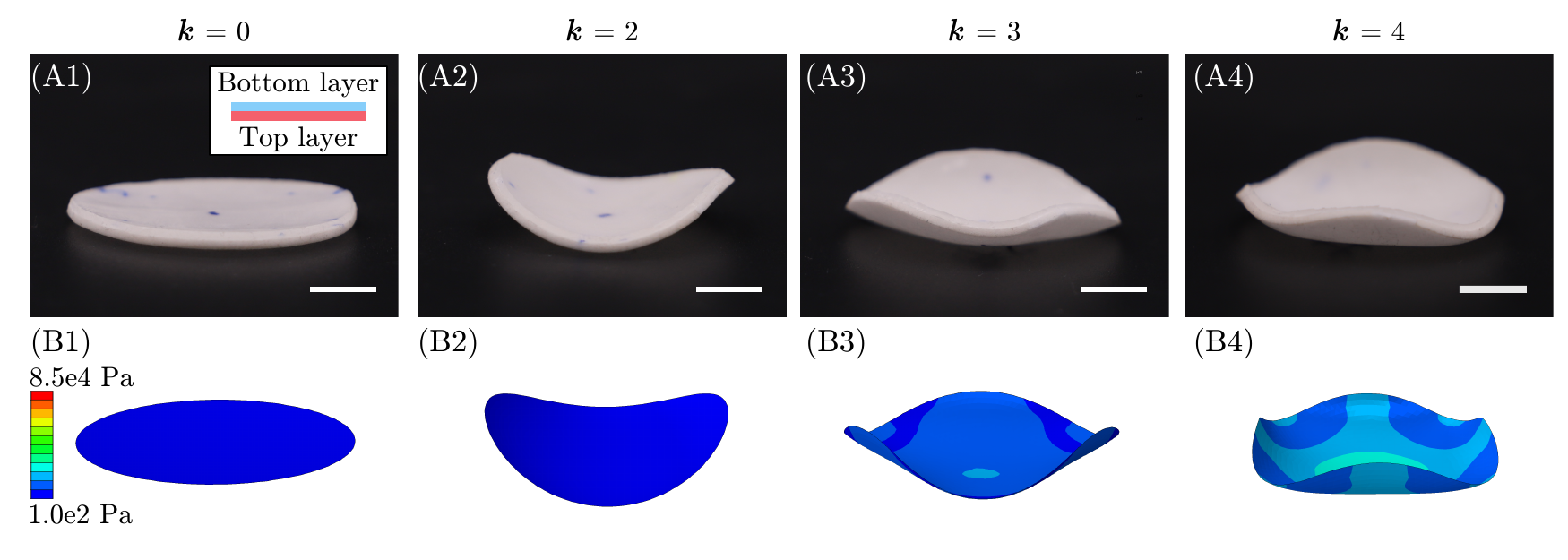}
    \caption{
    Free buckling shapes of soft composites. (A) Experimental images and (B) snapshots from finite element simulations at different values of prestretch: (1) $\lambda$ = 1.015, (2) $\lambda$ = 1.064, (3) $\lambda$ = 1.108, and (4) $\lambda$ = 1.167 to obtain different number of waves ($k=0, 2, 3$, and $4$, respectively) on the outer edge. Scalebar: 1 cm.
    For experimental parameters see Section~\ref{sec:parameters}.
    } 
    \label{fig:circularComposites}
\end{figure}

Our investigation starts with the setup described in Fig.~\ref{fig:overview}. A bottom layer of circular shape is first stretched by a prescribed amount of prestretch ($\lambda=1$ corresponds to stress-free configuration). A stress-free circular top layer is affixed onto the bottom layer. Physical parameters are described in \textbf{Materials and Methods.} Once the composite structure is released from the experimental setup of Fig.~\ref{fig:exptSetup}, a variety of shapes emerge depending on prestretch, $\lambda$. In Fig.~\ref{fig:circularComposites}A, a series of four shapes of the experimental composites are presented. These shapes -- that only differ by $\lambda$ -- are qualitatively different. The number of wrinkles on the outer edge of the composites, represented by the $k$ number in the figure, increases from $k=0$ in Fig.~\ref{fig:circularComposites}A1 to $k=4$ in Fig.~\ref{fig:circularComposites}A4. Fig.~\ref{fig:circularComposites}B shows finite element (FE) simulations that can also capture the qualitatively distinct shapes. While developing FE simulations, we realized the existence of different branches in the solution space and a propensity for the FE method to reach a local energy minimum instead of the global. This exposes a challenge in numerical simulation of such structures and motivates us to fundamentally understand the problem through a theoretical lens. The next section outlines the theory that was developed to explain the observations in Fig.~\ref{fig:circularComposites}.

\begin{figure}[h!]
\centering
   \includegraphics[width=0.90\textwidth]{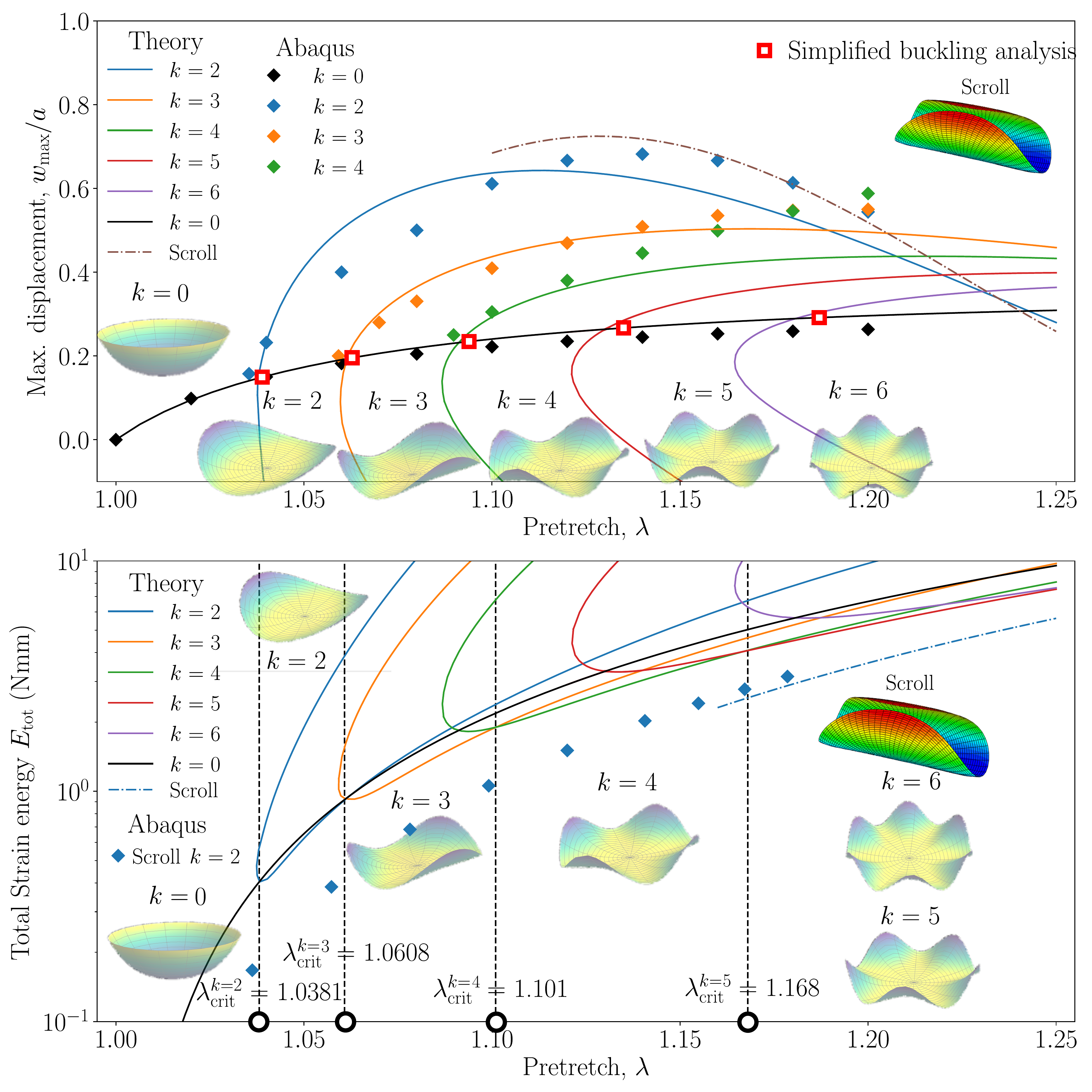}
   \caption{
   Theory with branches.
   The bottom figure plots the $k$ value as a function of prestretch, $\lambda$. For a given geometric and material properties there exists a critical $k$ after which the only possible (minimal energy) solution corresponds to the scroll-mode (i.e., there exists a max $k$).
   }
   \label{fig:fig_theory}
\end{figure}

\section*{Composite plate theory}

We assume plane stress and Kirchhoff-Love (KL) kinematic assumption on total in-plane displacement $\doubleunderline{u}(r,\theta,z) = \doubleunderline{u}^0(r,\theta) - z \nabla w(r,\theta)$ as the circular laminated composite plates used in experiments are thin. Here, $r$, $\theta$, $z$ are coordinates of the plate in radial, circular and direction across the thickness, $\doubleunderline{u}^0$ and $w$ are the displacement vector of the mid-surface and its vertical displacement, respectively. 
Furthermore, we use the Green-Lagrange strain tensor under F\"{o}ppl-von Karman (FvK) kinematic assumptions, which further simplifies the theory and limits it to small strains and moderate rotations, 
\begin{equation}
\doubleunderline{E}_\text{FvK} = \doubleunderline{E}_\text{FvK}^0 - z \nabla \otimes \nabla w,
\end{equation}
where $2 \doubleunderline{E}_\text{FvK}^0 = \nabla \doubleunderline{u}^0 + \left ( \nabla \doubleunderline{u}^0 \right )^\mathrm{T} + \nabla w \otimes \nabla w$ is the GL strain tensor of the mid-surface under FvK assumptions. In these expressions, $\nabla \otimes \nabla w$ is the curvature tensor, which will be later denoted by $\doubleunderline{\kappa}$ and $\nabla w \otimes \nabla w$ 
is the only nonlinear term remaining from the complete GL strain tensor.
For the material we assume that it is compliant with St. Venant-Kirchhoff elastic material model,
\begin{equation}\label{eqn:constitutive_equation}
	\doubleunderline{\sigma}_{K} = \frac{E_K}{1-\nu_K^2}\left( 
	(1-\nu_K) \left(\doubleunderline{E}_\text{FvK} - \doubleunderline{S}_{K} \right) + \nu_K \tr \left(\doubleunderline{E}_\text{FvK} - \doubleunderline{S}_{K} \right) \doubleunderline{I}
	\right),
\end{equation}
where $E_K$, $\nu_K$, $\doubleunderline{S}_{K}$ are Young modulus, Poisson ratio and prestrain tensor in $K$-th layer in the composite. As in experiment, we apply $\doubleunderline{S}_{K} = - \doubleunderline{I} (\lambda - 1)$, where $\lambda$ and $\doubleunderline{I}$ are the applied prestretch and a unity matrix, respectively.

A membrane force and bending moment (per unit length) tensors are as usually defined by $\doubleunderline{N} = \int_h \doubleunderline{\sigma} dz$ and $ \doubleunderline{M} = \int_h \doubleunderline{\sigma} z dz$, respectively, where $h$ is the overall thickness of the composite plate. 
A coupling between between $\doubleunderline{N}$, $\doubleunderline{M}$ as functions of $\doubleunderline{\kappa}$, $\doubleunderline{E}_{\text{FvK}}^0$ and the applied prestretch $\lambda$ is obtained
\begin{equation}\label{eqn:NMsistem}
    \begin{aligned}
        \doubleunderline{N} = &\; \doubleunderline{\overline{\alpha}}
         \doubleunderline{E}_{\text{FvK}}^0 + 
        \doubleunderline{\overline{\beta}} \doubleunderline{\kappa} + \doubleunderline{\overline{\gamma}}(\lambda),\\
        \doubleunderline{M} = & \doubleunderline{\tilde{\alpha}} \doubleunderline{E}_{\text{FvK}}^0 +
        \doubleunderline{\tilde{\beta}} \doubleunderline{\kappa} + \doubleunderline{\tilde{\gamma}}(\lambda).
    \end{aligned}
\end{equation}
Furthermore, expressing $\doubleunderline{E}_{\text{FvK}}^0$ from \eqref{eqn:NMsistem}$_1$ and inserting it into \eqref{eqn:NMsistem}$_2$ yields expression for $\doubleunderline{M}$ as a function of $\doubleunderline{\kappa}$ and $\doubleunderline{N}$,
\begin{equation}\label{eqn:Main_compact_form_short}
    \begin{aligned}
        \doubleunderline{M} = \doubleunderline{{e}} \doubleunderline{N} -
        \doubleunderline{\overline{D}} \doubleunderline{\kappa} + \doubleunderline{{\gamma}}_0.
    \end{aligned}
\end{equation}
Here, symbols $\doubleunderline{\overline{\alpha}}$, $\doubleunderline{\overline{\beta}}$, $\doubleunderline{\overline{\gamma}}$,
$\doubleunderline{\tilde{\alpha}}$, $\doubleunderline{\tilde{\beta}}$, $\doubleunderline{\tilde{\gamma}}$, $\doubleunderline{e}$, $\doubleunderline{\overline{D}}$, $\doubleunderline{\gamma}_0$ denote auxiliary functions that are defined later in this section, see Eqs. (\ref{eqn:stiffnessesTensors}) and (\ref{eqn:stiffnesses}).

Let us define a stress function $\mathcal{F}$ such that 
\begin{equation}\label{eqn:nablaFN}
    \tilde{\nabla} \tilde{\nabla} \mathcal{F}  := \Delta \mathcal{F}\doubleunderline{I} - \nabla\otimes\nabla \mathcal{F} = \doubleunderline{N}
\end{equation}
and that satisfies the in-plane equilibrium 
$\nabla \cdot \doubleunderline{N} = \doubleunderline{0}$. 
Inserting Eqs. (\ref{eqn:Main_compact_form_short}) and (\ref{eqn:nablaFN}) into the out-of-plane equilibrium of the plate $\nabla \cdot (\nabla \cdot \doubleunderline{M}) + \nabla \cdot \left ( \doubleunderline{N} \cdot \nabla w\right) = 0$ its alternative form is obtained  
\begin{equation}\label{eqn:Main_eqnsF}
    \overline{D} \Delta^2 w  - \nabla \cdot \left (\tilde{\nabla} \tilde{\nabla} \mathcal{F}  \nabla w\right)  =  0.
\end{equation}
Similarly, an alternative form of the compatibility conditions $\nabla \times \left( \nabla \times \doubleunderline{E}^0_{\text{FvK}} \right)^{\mathrm{T}} = \boldsymbol{0}$ can be written as
\begin{equation}\label{eqn:komp}
    \Delta^2 \mathcal{F} + \frac{\overline{\alpha} (1-\overline{\nu}_A^2)}{2} \left[ w , w\right] = 0,
\end{equation}
where $\left[\phantom{.},\phantom{.}\right]$ is the Monge-Ampere operator.


\begin{equation}\label{eqn:stiffnesses}
    \begin{aligned}
        \tilde{\alpha} & =  \left (  \frac{E_2}{1-\nu_2^2} e_2 h_2 - \frac{E_1}{1-\nu_1^2} e_1 h_1 \right )\\
        \tilde{\beta} & =  - \left (  \frac{E_2}{1-\nu_2^2} \left (\frac{h_2^3}{12}  + h_2 e_2^2\right ) + \frac{E_1}{1-\nu_1^2} \left (\frac{h_1^3}{12}   + h_1 e_1^2 \right ) \right )\\
        \tilde{\gamma} & =  - \left (  \frac{E_2}{1-\nu_2}  h_2 e_2 \eta_2 - \frac{E_1}{1-\nu_1}  h_1 e_1 \eta_1       \right )\\
        \tilde{\nu}_A & =  \frac{1}{\tilde{\alpha}}
                \left (  \frac{E_2 \nu_2}{1-\nu_2^2} e_2 h_2 - \frac{E_1 \nu_1}{1-\nu_1^2} e_1 h_1 \right )\\
        \tilde{\nu}_B & =  \frac{-1}{\tilde{\beta}} \left (  \frac{E_2 \nu_2}{1-\nu_2^2} \left (\frac{h_2^3}{12}  + h_2 e_2^2\right ) + \frac{E_1 \nu_1}{1-\nu_1^2} \left (\frac{h_1^3}{12}   + h_1 e_1^2 \right ) \right )\\
        \overline{\alpha} & =   \left (  \frac{E_2}{1-\nu_2^2} h_2 + \frac{E_1}{1-\nu_1^2}  h_1 \right )\\
        \overline{\beta} & =   - \tilde{\alpha}\\
        \overline{\gamma} & =  - \left (  \frac{E_2}{1-\nu_2}  h_2  \eta_2 + \frac{E_1}{1-\nu_1}  h_1  \eta_1       \right )\\
        \overline{\nu}_A & =  \frac{1}{\overline{\alpha}}
                \left (  \frac{E_2 \nu_2}{1-\nu_2^2}  h_2 + \frac{E_1 \nu_1}{1-\nu_1^2}  h_1 \right )\\
        \overline{\nu}_B & =  \tilde{\nu}_A 
    \end{aligned}
\end{equation}

\begin{equation}\label{eqn:stiffnessesTensors}
    \begin{aligned}
        \doubleunderline{\overline{\alpha}} & = 
        \overline{\alpha}
        \begin{bmatrix}
        1  & \overline{\nu}_A & 0\\
         \overline{\nu}_A  & 1  & 0 \\
        0 & 0 &  2(1 - \overline{\nu}_A)
        \end{bmatrix}\\
        \doubleunderline{\overline{\beta}} & = 
        \overline{\beta}
        \begin{bmatrix}
        1  & \overline{\nu}_B & 0\\
         \overline{\nu}_B  & 1  & 0 \\
        0 & 0 &  2(1 - \overline{\nu}_B)
        \end{bmatrix}\\
        \overline{\doubleunderline{\gamma}} & = \doubleunderline{I}\; \overline{\gamma}\\
        \doubleunderline{\tilde{\alpha}} & = 
        \tilde{\alpha}
        \begin{bmatrix}
        1  & \tilde{\nu}_A & 0\\
         \tilde{\nu}_A  & 1  & 0 \\
        0 & 0 &  2(1 - \tilde{\nu}_A)
        \end{bmatrix}\\
        \doubleunderline{\tilde{\beta}} & = 
        \tilde{\beta}
        \begin{bmatrix}
        1  & \tilde{\nu}_B & 0\\
         \tilde{\nu}_B  & 1  & 0 \\
        0 & 0 &  2(1 - \tilde{\nu}_B)
        \end{bmatrix}\\
        \tilde{\doubleunderline{\gamma}} & = \doubleunderline{I}\; \tilde{\gamma}.
    \end{aligned}
\end{equation}


\subsection{Uncut symmetric solution}\label{subsec:sim_res}
Following experimental results shown in Fig. \ref{fig:circularComposites} we first seek an axi-symmetric solution, which we refer to as a ``cup''. The theory suggests that after the prestretched layer is adhered to a nonprestretched one the membrane stresses are equlibrated in both layers after the release, but due to the coupling between membrane stresses in the mid-surface of each layer, bending moments are induced. 
These lead to the shortening and compressive (circular) membrane forces in the circumference of the two-layered circular composite plate, most prominently towards the edge. By plane equilibrium, tensile membrane forces are induced towards the center of the circular plate.

Due to the symmetry of the problem, the order of both PDEs (\ref{eqn:Main_eqnsF}) and (\ref{eqn:komp})  that describe the deformation of the composite plate can be reduced by substitution $\varphi = w_{,r}$ and $\phi = \mathcal{F}_{,r}$. After integration we obtain
\begin{equation}\label{eqn:Main_2nd_order_eq}
\begin{aligned}
    \overline{D}  r \left( \Delta \varphi - \frac{\varphi}{r^2} \right)
    -\phi \varphi &= 0 \\
    \frac{1}{\overline{\alpha}(1-\overline{\nu}_A^2)} r \left ( \Delta \phi - \frac{\phi}{r^2} \right) + \frac{\varphi^2}{2} &= 0.
 \end{aligned}
\end{equation}
We approximate $\varphi$ with the model function $\varphi(r) = C_1 r^n$, to obtain $\phi(r) = C_1^2  r/a ( 1 - ( r/a) ^{2 n}) / (8 \Lambda  n (n+1))$ from \eqref{eqn:Main_2nd_order_eq}$_2$.
From the boundary conditions $M_{rr} = 0$ and $N_{rr} = 0$ at the edge, we obtain $C_1 =  \gamma_0 a / (\overline{D}(\nu + n))$. Here, we took $\nu = \tilde{\nu}_A \doteq \tilde{\nu}_B \doteq \overline{\nu}_A \doteq \overline{\nu}_B$ for the sake of simplicity. The exponent $n$ can be determined by applying orthogonality condition on \eqref{eqn:Main_2nd_order_eq}$_1$,
$\int_{0}^{a} \left ( \overline{D} r \left ( \Delta \varphi - {\varphi}/{r^2} \right ) - 	\phi    \varphi \right ) {\partial \varphi}/{\partial n}   dr =  0$.
This condition yields an implicit relation between $n$, the dimensionless parameter $\chi$ and $\nu$ in the following form $3 (2 n+1)^2 \chi -4 \left(n^2-1\right) \left(8 n^2+18 n+9\right)^2 (\nu +n)^2 = 0$, where $\chi = B^2 a/(\Lambda \overline{D})$, $B = \gamma_0 a / \overline{D}$ and $\Lambda = 1/((1-\overline{\nu}_A^2)\overline{\alpha} a)$, while $\gamma_0$ is the moment due to prestretch. The explicit expression for $n(\chi(\lambda))$ is solved numerically.

From experiments and simulations described earlier we learn that by increasing the prestrecth, the axisymmetric cup structure losses its stability and wrinkles into a $k$-fold axisymmetric structure when the circular forces exceed some critical.

\subsection{\emph{k}-fold axisymmetric postbuckling}
We seek the $k$-fold axisymmetric solution with the following two model functions $w(r,\theta) = \overline{w}(r) + \tilde{w}(r,\theta)$ and $\mathcal{F}(r,\theta)  = \overline{\mathcal{F}}(r) + \tilde{\mathcal{F}}(r,\theta)$, where
$\overline{w}(r) = \int \varphi(r) dr$ and $\overline{\mathcal{F}}(r) = \int \phi(r) dr$ are the axisymmetric solutions from before and $\tilde{w}(r,\theta) = f(r) \cos k \theta$ and $\tilde{\mathcal{F}}(r,\theta) =  g(r) \cos k \theta$ are the $k$-fold axisymmetric eins\"{a}tze. Plugging both model functions into Eqs. (\ref{eqn:Main_eqnsF}) and (\ref{eqn:komp}) yields
\begin{equation}\label{eq:Main_MDMVTheory}
	\begin{aligned}
		\overline{D} \Delta^2 \tilde{w} -  \doubleunderline{\kappa}_0 : (\nabla \otimes \nabla{\tilde{\mathcal{F}}})
		- \doubleunderline{N}_0 : (\nabla \otimes \nabla \tilde{w} )
		-  \left[ \tilde{\mathcal{F}} , \tilde{w}  \right] & = 0, \\
		\frac{\Delta^2 \tilde{\mathcal{F}} }{\overline{\alpha}(1-\overline{\nu}_A^2)}  +
		\doubleunderline{\kappa}_0 : (\nabla \otimes \nabla \tilde{w})  +
		\frac{1}{2} \left[ \tilde{w}  , \tilde{w}  \right]
		& = 0.
	\end{aligned}
\end{equation}
Here, $\doubleunderline{\kappa}_0 = \Delta \overline{w} \doubleunderline{I} - \nabla \otimes \nabla \overline{w}$ and $\doubleunderline{N}_0$ are modified curvature tensor and membrane force tensor of the symmetric solution, respectively. 

We observe that for smaller prestretches $\lambda$ the first most energetically favourable (stable) wrinkling mode is 2-fold symmetric (Pringles chip-like), followed by 3-fold, 4-fold etc.  symmetric solutions, when the prestretch is increased.
Fig. \ref{fig:fig_theory}, where the occuring deformation modes and their strain energies are displayed. With each following mode the wrinkles are more and more localized at the edge, because with deeper and deeper spherical cap, the membrane compressive stresses are more and more localized at the edge (see Fig. \ref{fig:InnerMembraneForces}) and then released through wrinkling. Because wrinkles with large $k$-fold symmetry are more localized at the edge, they are energetically more favorable, because they release more of the membrane energy, which is localized at the edge.

Another reason why the higher deformation modes $k$ are promoted with larger prestretches $\lambda$ is that the increasing curvature of the system (spherical cup) provides geometric rigidity, which acts analogously to an elastic substrate. This is evident if we linearize \eqref{eq:Main_MDMVTheory}$_2$ and insert it into \eqref{eq:Main_MDMVTheory}$_1$. We obtain an equation of the form $\tilde{D} \Delta^2 \tilde{w} - \doubleunderline{N_0} : \nabla \otimes \nabla \tilde{w} + \alpha_{0} (\kappa_0:\kappa_0) \tilde{w} = 0$, which is analogous to the equation that governs wrinkling of thin compressed films on compliant substrates. The therm $\alpha_{0} (\kappa_0:\kappa_0) \tilde{w}$ penalizes buckling modes with large amplitude to wavelength ratios, therefore buckling modes with larger $k$ are energetically preferable.

\subsection{Deep postbuckling}
When the stretching $\lambda$ of the membrane is large we observe that $k$-fold symmetric solution is no longer energetically favorable. Instead, we detect experimentally and computationally (using Abaqus) that cylindrical bending solution is dominant. We refer to this solution as a ``scroll'' solution. This reduces Eq.~\eqref{eqn:Main_eqnsF} to 
$\kappa_{11,11}  =  0$, as only $\kappa_{11} \neq 0$ and $\kappa_{22} = \kappa_{12} = 0$, while Eq.~\eqref{eqn:komp} is automatically satisfied.
Assuming stress free edges, we obtain $\doubleunderline{N} = \doubleunderline{0}$ and $\kappa_{11} = \gamma_0/\overline{D}$. The strain energy is calculated as $\mathcal{E} = \pi a^2 \gamma_0^2/(2\overline{D})$.

The $k$-fold symmetries do not increase indefinitely. At a certain point the scroll solution (which is similar to the 2-fold symmetry) offers the largest relaxation of the membrane stresses. Its membrane stresses are released entirely, however the price for this is a bit larger bending strain energy, than in the $k$-fold symmetric cases, therefore it is not preferable with smaller $\lambda$.

\begin{figure*}[h!]
\centering
    \includegraphics[width=1\textwidth]{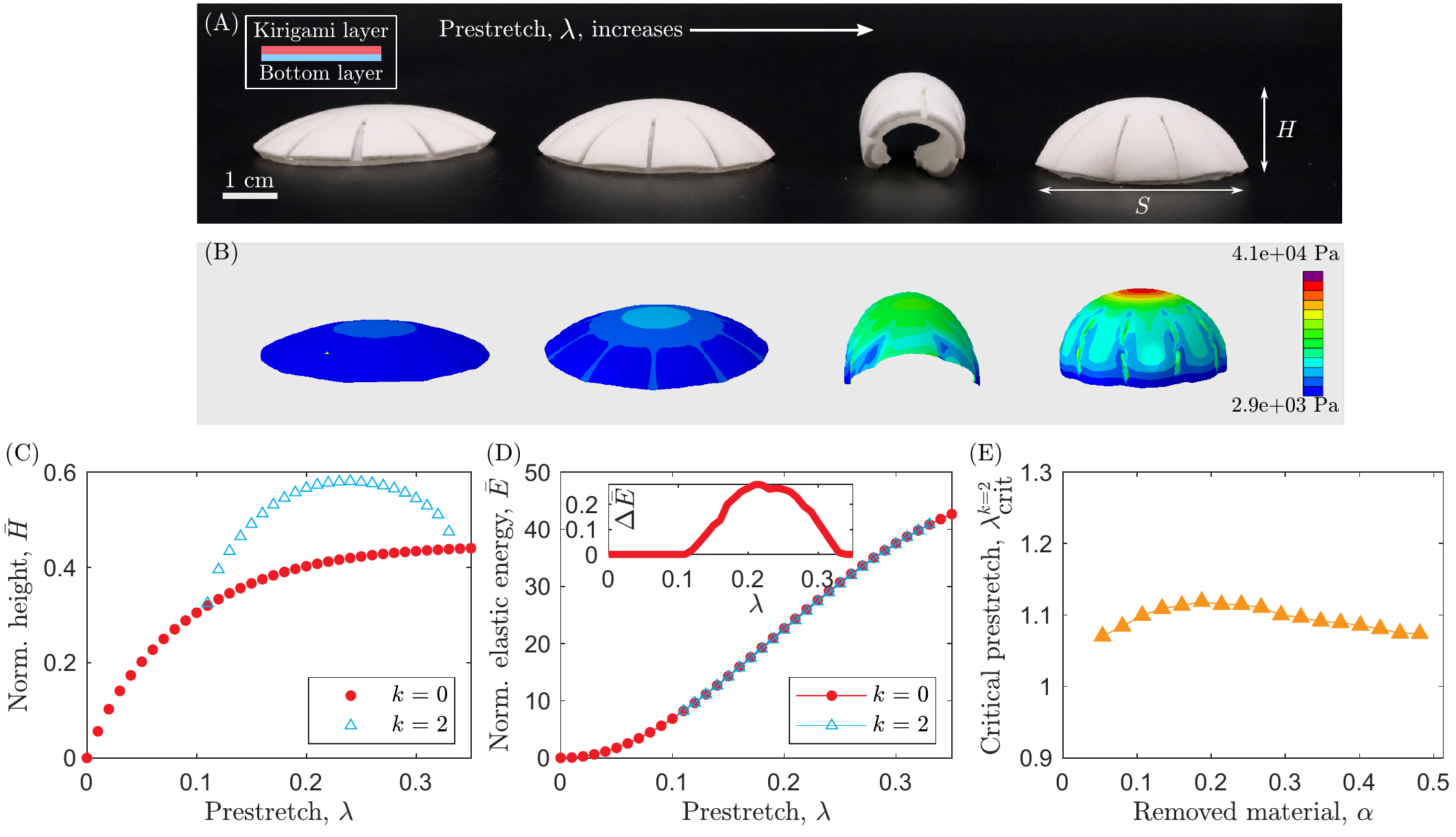}
    \caption{
    Soft kirigami composites. (A) Experimental images and (B) snapshots from finite element simulations at different values of prestretch: (1) $\lambda = 1.05$, (2) $\lambda=1.15$, (3) $\lambda=1.2$, and (4) $\lambda=1.25$ to obtain the preprogrammed shapes. It turns out that a (quasi)axisymmetrical shape is not energetically favorable for $1.15 \lesssim \lambda \lesssim 1.25$ for this particular set of material and geometric parameters.
    (C) Normalized height, $\bar H = H/R_o$, as a function of prestretch, $\lambda$, where $R_o$ is the radius of the kirigami layer.
    (D) Normalized strain energy, $\bar E$, as a function of $\lambda$ at two different mode shapes: $k=0$ and $k=2$.
    The strain energy has been normalized by the characteristic bending energy, as described in the text. The inset shows the difference in normalized strain energy between the two ($\Delta \bar E = \bar E_{k=0} - \bar E_{k=2}$, where the subscript indicates the mode number) as a function of $\lambda$.
    (E) Critical prestretch, $\lambda^{k=2}_{\textrm{crit}}$, as a function of the portion of the removed material, $\alpha$ ($\alpha=1$ means the entire layer has been removed).
    }
    \label{fig:kirigamiComposites}
\end{figure*}

\section*{Soft Kirigami Composites}

Informed by the nonlinear nature of the problem and the existence of multiple branches in the previous section, we essentially need to find out a way of reaching the hemispherical cap solution. Therefore, we need to achieve at least ``quasi-axisymmetric", if not an axi-symmetric ($k=0$) solution at at larger values of prestretch. A opposed to the 1D example, where applying enough prestretch enables us to produce a circle, here on the circular plate circular compressive forces \cite{stein2019buckling, Holmes2019COCIS, Pezzulla2015SM} occuring due to the change of circumference during bending, prevent that the circular plate would bend into a hemisphere, as the $k$-fold wrinkling inastability or scroll instability occurs at the edge \cite{Hadrien2017SM, Holmes2019COCIS,Pezzulla2015SM}.

The first solution that comes to mind \cite{moshe2019kirigami, tang2019programmable, cui2018origami} is therefore to cut the top layer along a prescribed path, whereby forming a so called ``soft kirigami composite" as a way of releasing the compressive force from certain areas of the composite and partially remove the wrinkled nature of the outer edge. 
A heuristic shape for the top layer was chosen as shown in Fig.~\ref{fig:overview}B. 
Physical parameters are presented in \textbf{Materials and Methods}. Fig.~\ref{fig:kirigamiComposites}A and B present snapshots of the structural shapes at different values of prestretch from experiments and simulations, respectively.

If we remove a circumferential fraction of material $\alpha$, which is divided between $n_c$ pieces, we enforce $n_c$-fold symmetry, which is unfortunatelly only  ``quasi-axisymmetric" and not axi-symmetric $k=0$.  This is because, even at small prestretches $\lambda$, but even more so at the prestretches larger than the critical prestretch for $n_c$-fold buckling $\lambda > \lambda_\text{crit}^{k=n_c}$ the inhomoegnity in the circular direction causes additional bending of the ``pizza slices''  in the circular direction. Therefore to obtain a solution that is as axi-symmetric as possible a large number $n_c \to \infty$ pieces of composite have to be removed. 
However this might not be the best idea since both, the number of cuts $n_c$ and the choice of the cut out ratio $\alpha$ affect weather the desired $n_c$-fold symmetric hemisphere is stable or weather it looses stability and buckles into an undesired low $k$-fold symmetric deformation mode.
Therefore we are searching for the appropriate $\alpha$ and $n_c$  to ensure that:
 1.) That the $n_c$ -fold symmetric hemisphere is stable.
 2.) That it is symmetric enough, since 2,3,4,5-fold symmetric shapes are not symmetric enough for our liking.

We employed an empirical trial and error approach to find a suitable set of parameters for a shape close to a hemispherical cap in Fig.~\ref{fig:kirigamiComposites}A-B at $\lambda =1.25$. Fig.~\ref{fig:kirigamiComposites}C and D show the height, $H$, of the kirigami composite and the elastic energy, $E$, as functions of prestretch, $\lambda$. The height has been normalized by the radius of the substrate ($R=2$ cm) so that $\bar H = H / R$. The strain energy has been normalized by the characteristic elastic energy of the composite as discussed in \textbf{Materials and Methods.} The difference in energy between two different modes ($k=0$ and $k=2$) is too small to be visually observable, therefore, Fig.~\ref{fig:kirigamiComposites} includes an inset showing the normalized difference in strain energy between the two modes: $\Delta \bar E = \bar E_{k=0} - \bar E_{k=2}$. Note that $\Delta \bar E$ is positive between $1.15 \lesssim \lambda \lesssim 1.30$; i.e., $k=2$ is energetically favorable compared with $k=0$ shape.

This critical value of prestretch, $\lambda_\textrm{crit}^{k=2}$, at which $k=2$ shape appears from a $k=0$ shape depends on the amount of removed material in the Kirigami layer, $\alpha$, among other parameters. $\alpha=1$ means the entire layer has been removed, whereas $\alpha=0$ corresponds to a circular shape with no Kirigami cuts at all. Fig.~\ref{fig:kirigamiComposites}E shows the critical prestretch, $\lambda_\textrm{crit}^{k=2}$, as a function of $\alpha$ from finite element simulations and theoretical analysis.

\section{Materials and Methods}

\subsection*{Physical Parameters}
A commercially available acrylic adhesive (3M VHB tape) is used as the bottom and kirigami layers. Mooney-Rivlin model is used in the finite element simulations of Fig.~\ref{fig:circularComposites} and Fig.~\ref{fig:kirigamiComposites}. The strain energy per unit of reference volume in this model is
\begin{equation}
    U = C_{1} \left( \bar I_1 - 3 \right) + C_{2} \left( \bar I_2 - 3 \right) + D_1 \left( J - 1 \right)^2,
\end{equation}
where $C_1, C_2$, and $D_1$ are material parameters, $\bar I_1$ and $\bar I_2$ are the first and second deviatoric strain invariants, and $J$ is the elastic volume ratio. The bottom layer has a thickness of $h=1.0$~mm and its material parameters are $C_1 =4.84$~kPa, $C_2=14.536$~kPa, and $D_1 = 0.96\times10^{3}$~kPa. The kirigami/top layer is $h=1.1$~mm thick with the following material parameters:
$C_1 = 4.00$~kPa, $C_2 = 78.7152$~kPa, and $D_1=4.1082\times10^{3}$~kPa.

In order to normalize the elastic strain energy, we formulate a  characteristic elastic energy of the composites: $E^* = Y h^3$, where $Y = 2 G (1 + \nu)$ is the effective Young's modulus of the substrate, $h$ is the thickness of the substrate, and $\nu=\frac{3K-2G}{6K+2G}$, $G = 2(C_1 + C_2)$, and $K=2/D1.$

For the soft kirigami composites in Fig.~\ref{fig:kirigamiComposites}, the number of cuts is  $n_c=10$ and the ratio of the inner radius to the outer radius is $\bar r = 0.4$ and $d\theta=0.1$ radians, i.e., $1$ out of $2\pi$ radian was removed. This corresponds to $\alpha=0.13$.

\end{document}